\documentclass[11pt]{article}
\usepackage{graphicx}

\title{\bf Anisotropic flows from colour strings: Monte-Carlo simulations}
\author{ M.A.Braun$^{a,b}$, C.Pajares$^a$, V.V.Vechernin$^b$ \\
 $^a$ University of Santiago de Compostela, Spain,\\
$^b$ S.Petersburg State University, Russia
}

\begin{document}
\maketitle
\def\beq{\begin{equation}}
\def\eeq{\end{equation}}
\def\bp{{\bf p}}

{\bf Abstract}
By direct Monte-Carlo simulations it is  shown that the anisotropic
flows can be successfully described in the colour string picture with
fusion and percolation provided  anisotropy of
particle emission from the fused string is taken into account. Quenching
of produced particles in the strong
colour field of the string is the basic mechanism for this anisotropy.
The concrete realization of this mechanism is borrowed from the
QED. Due to dependence of this mechanism on the external field strength
the found flows grow with energy, with values for $v_2$ at LHC energies
greater by ~15\% than at RHIC energies.

\section{Introduction}
The study of anisotropic flows in the spectra of particles produced
in high-energy heavy-ion collisions proves to be very informative
as to the dynamics of the underlying emission mechanisms.
Observed experimental values for the anisotropic flow coefficients $v_n$
impose strong restrictions on the models which try to describe the data.
Of special importance are the coefficients with odd values of $n$, which
disappear on the average and can only be observed on the event-by-event
basis. They carry information about the details of the evolution of
the hot nuclear matter in the overlap and its fluctuations
during the collision.
These fluctuations in principle may come from both initial conditions
for the nuclear matter evolution  and
dynamics of the successive evolution itself. Simple phenomenological
initial conditions based on  fluctuations in the distribution of
participant nucleons followed by the hydrodynamical evolution
allow to describe the third harmonic coefficient
$v_3$ quite successfully ~\cite{glau1, glau2,glau3,glau4}. Remarkably
less phenomenological initial conditions which follow from the
preasymptotic evolution of emitted gluons in the Colour Glass
Condensate approach seriously underestimate $v_3$
~\cite{glau1,glau4} unless this evolution is treated in a rigorous
non-perturbative manner and the subsequent evolution has a
non-vanishing viscosity ~\cite{venugo}

In ~\cite{brapaj1} we  drew attention to the fact that the elliptic flow
could also be naturally explained in the colour string approach,
with string fusion and percolation \cite{armesto,
braun1,braun2,braun3}, quite successful in the description
of particle production and correlations
in the soft part of spectra \cite{braun4,DDD,cunqueiro}.
In this approach partons emitted at some point
have to pass a certain length before they appear outside
and are observed. On their way they cross a strong colour field
inside the strings and emit gluons, so that
their energy diminishes. As a result, the observed particle
energy turns out to be smaller than at the moment of its creation and
this energy quenching depends on the path length passed inside
the nuclear matter and so on the direction of its
transverse momentum $p$. Very crude estimates made in ~\cite{brapaj1}
confirmed  that a sizable elliptic flow followed from this
mechanism. Its centrality and transverse momentum dependence qualitatively
agreed with the behaviour of the RHIC data \cite{adler,adare,alver,voloshin}.
These results also supported the ones obtained in a similar framework
using different simplified methods ~\cite{bautista1,bautista2,bautista3}.

It was stressed in ~\cite{brapaj1} that any trustworthy and
comprehensive quantitative results could only be achieved
in more elaborate studies based on  detailed
Monte-Carlo simulations. This is especially true for odd flow
harmonics which, as mentioned, can only be studied on the event-by-event
basis, with all sources of fluctuations taken into account.
Calculations along these
lines constitute the subject of this paper.
On the general they confirm crude predictions made in
~\cite{brapaj1} and also allow for a comprehensive study of azimuthal
anisotropy in particle production and thus determination of
higher anisotropic coefficients $v_n$ with $n>2$.
Note that the colour string approach combines both the initial conditions
created in the course of collisions and the subsequent evolution of the
nuclear matter in the collision zone modeled by the fusion of produced
strings and their final decay into observed hadrons.
Fluctuations may be present at all stages of this process. Our calculations
show that the decisive role for the successful description of both
$v_2$ and $v_3$ is played by the fluctuations in the initial geometry of
the collision plus those in the string fusion. Fluctuations in the
final production of observed particles play a minor role. Our model
has a single new parameter, the quenching coefficient, which characterizes
energy loss of  the produced parton in the string matter. We adjust
this parameter to agree with coefficient $v_2$ observed in mid-central
Au+Au collisions at 200 GeV and integrated over the transverse momenta.
With thus adjusted parameter we are able to explain $v_2$ at all
centralities and transverse momenta for collisions at 62.4 GeV,
200 GeV and 2.76 TeV. The third harmonic
$v_3$ comes out a little smaller than the average observed values. However
the event-by-event fluctuations in $v_3$ (as well as in $v_2$) are
calculated to be
quite large so that the observed values are found to be within the
calculated ones when these fluctuations are taken into account.

The structure of the paper is as follows. In the next section we briefly
discuss our model and
also present our quenching mechanism
borrowed from QED. In Section 3 we describe our Monte-Carlo procedure
to study the anisotropic flow on the event-by-event basis.
Section 4. presents our numerical results. In Conclusion we draw some
lessons from our study and point out some possible further refinements.

\section{The model}
One needs
to combine two ingredients to have anisotropic  flows. First, string
fusion has to generate clusters which are azimuthally asymmetric and
emit particles anisotropically. But by itself this will not produce
any elliptic flow unless the distribution of these clusters in the
transverse plane is also azimuthally asymmetric. This latter phenomenon
can only occur if the clusters are large enough to feel the asymmetric form
of the overlapping region in the collision. Such clusters arise in the
process of percolation of fused strings.

In our model it is assumed that at the moment of the collision
color strings are stretched between partons of the colliding nuclei.
Since they are many and so overlap in the transverse space, they
fuse and percolate to form macroscopic clusters at some critical string
density $\rho=\rho_c$
where
\beq
\rho=\frac{N\Omega_0}{\Omega},
\label{rho}
\eeq
 $\Omega_0$ is the transverse area of simple strings, $N$ is their
number and $\Omega$ is the nuclear overlap area.
 Starting from the moment of their formation strings
decay into particles (quark-antiquark pairs), which process we describe
using the well-known formalism for pair creation in a strong
electromagnetic field. According to this mechanism, in its simplest version,
the particle distribution at the moment of its production by the string is
\beq
P(p,\phi)=Ce^{-\frac{p_0^2}{T}}.
\label{prob}
\eeq
where $p_0$ is the particle initial transverse momentum,
$T$ is the string tension (up to an irrelevant numerical coeffcient) and
$C$ is the normalzation factor.
However $p_0$ is different from the observed particle momentum $p$ because
the particle has to pass through the  fused string area and emit gluons
on its way out. So in fact in Eq. (\ref{prob}) one has to consider $p_0$
as a function of $p$ and path length $l$ inside the nuclear overlap:
$p_0=f(p,l(\phi))$ where $\phi$ is the azimuthal angle. Note that Eq. (\ref{prob})
describes the spectra only at very soft $p_0$. To extend its validity to
higher momenta one may use the idea that the string tension fluctuates, which
transforms the Gaussian distribution into the thermal one ~\cite{bialas,deupaj}:
\beq
P(p,\phi)=Ce^{-\frac{p_0}{\sqrt{T/2}}}.
\label{probb}
\eeq

Radiative energy loss has been extensively studied for a parton passing
through the nucleus or quark-gluon plasma as a result of multiple collisions
with the medium scattering centers ~\cite{baier}. In our case the situation
is somewhat different: the created parton moves in the external gluon field
inside the string. In the crude approximation this field can be taken as
being constant and  orthogonal to the direction of the parton propagation.
In the same spirit
as taken for the mechanism of pair creation,
one may assume that the reaction force due to radiation is similar to the
one in the QED when a charged particle is moving in the external
electromagnetic field. This force causes a loss of energy which for an
ultra-relativistic particle is proportional to
[its momentum$\times$field]$^{2/3}$ ~\cite{niri}:
\beq
\frac{dp(x)}{dx}=-0.12e^2 \Big(eEp(x)\Big)^{2/3},
\label{vbb}
\eeq
where $E$ is the external electric field.
Eq. (\ref{vbb}) leads to the quenching formula
\beq
p_0(p,l)=p\Big(1+\kappa p^{-1/3}T^{2/3}l\Big)^3,
\label{quench1}
\eeq
where we identified $eE/\pi=T$ as the string tension. As mentioned in the Introduction,
the quenching coefficient $\kappa$ is adjusted to
give the experimental value for the coefficient $v_2$
in mid-central Au+Au collisions at 200 GeV, integrated over the
transverse momenta.

Of course the possibility to use electrodynamic formulas for
the chromodynamic case may raise certain doubts. However in
~\cite{mikhailov} it was found that at least in the $N=4$ SUSY
Yang-Mills case the loss of energy of a coloured charge moving
in the external chromodynamic field was given by essentially the same
expression as in the QED.

Note that from the moment of particle creation to the moment of its
passage through other strings a certain time elapses depending on the
distance and particle velocity. During this time strings decay and
the traveling particle will meet another string partially
decayed, with a smaller colour $Q$ than at the moment of its formation.
So one has to consider a non-static string distribution
with string colours evolving in time and gradually diminishing until strings
disappear altogether. The time scale of this
evolution is estimated to be  considerably greater  than time intervals characteristic
for partons traveling inside the string matter. However the effect of
string decay with time is noticeable and we take it into account in our
calculations.

To study the time evolution of strings we  again turn to the Schwinger
mechanism.  For it one has the probability
of pair creation in unit time and unit volume as ~\cite{nikishov}
\beq
\Gamma_{Vt}=\frac{1}{4\pi}T^2e^{-\frac{p_0^2}{T}},
\eeq
where again $T$ stands for $eE/\pi$ in QED. For a realistic string the volume
$V=SL_z$ where $S$ is the string transverse area and $L_z$ is
the longitudinal dimension of the string.
For the string of colour $Q$ is $T=QT_0$ where
$T_0$ is the string tension of the ordinary string with $Q=1$. The average
transverse momentum squared of the emitted quark-antiquark pair
$<p_0^2>$ is just $T$.To estimate $L_z$ we assume that the string emits a
pair when its energy
is equal to $2<p_0>=\sqrt{T}$, which gives $L_z=1/\sqrt{T}$, so that
we get the average probability in unit time
\beq
\Gamma_t=\frac{1}{4\pi}T^{3/2}S.
\eeq
The string colour diminishes by unity with each pair production.
So we find an
equation which describes the time evolution of the string colour $Q(t)$
\beq
\frac{dQ(t)}{dt}=-\alpha Q^{3/2}(t)
\eeq
with the solution
\beq
Q(t)=\frac{Q_0}{(1+\frac{1}{2}\alpha t \sqrt{Q_0})^2}
\label{qtime}
\eeq
where $Q_0$ is the initial colour at the moment of the string creation.
Coefficient $\alpha=T_0^{3/2}S/(2\pi)$ depends on the string transverse
area $S$. As will be explained in next section, we use the picture in which
the fused string is in fact modeled by a set of "ministrings" formed at
intersections of simple strings with the same area as the simple string, but
greater color. This gives $\alpha= 0.03$ 1/fm.
This value has been used in our calculations.
The average color of ministrings is of the order 2 - 3. So it changes
only by  30 -50 \% even when the emitted parton travels 5 fm of distance.
Still this effect is felt in the calculations. In fact it
practically does not change the results but changes the value
of the quenching coefficient $\kappa$, which in any case is to be adjusted,
as explained above. In this sense our results are practically independent
of the concrete choice of $\alpha$ in the reasonable interval of values.

\section{Monte-Carlo simulations}
\subsection{Generalities}
\begin{figure}
\hspace*{2.5 cm}
\begin{center}
\includegraphics[scale=0.85]{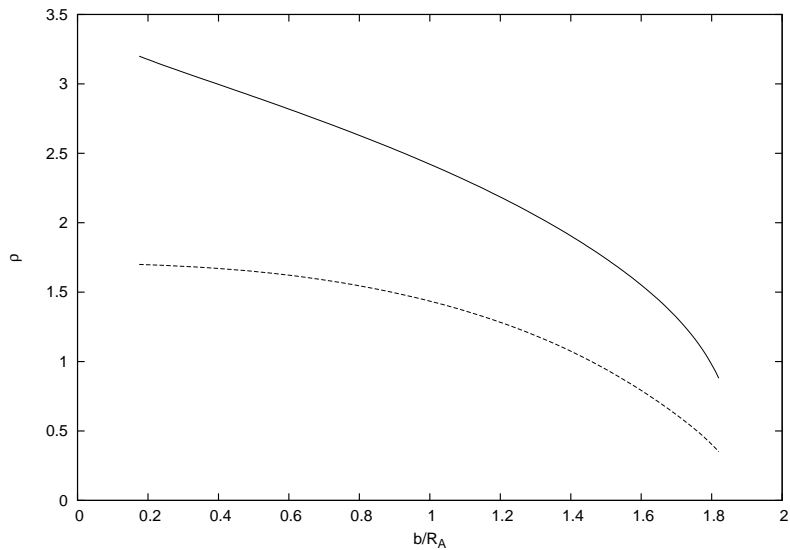}
\end{center}
\caption{Percolation parameter as a function of impact parameter
for Au-Au collisions at 62.4 (lower curve) and 200 GeV
borrowed from \cite{tarn}}
\label{fig1}
\end{figure}

In principle Monte-Carlo simulations of the quenching in the fusing string
scenario seem to be straightforward. One models  strings in the
nuclear overlap by discs of a given radius $r_0$.  In an event
$N$ discs are assumed to be distributed in the transverse plane
in agreement with the distribution of the participant ("wounded")
nucleons, given by product of the nuclear profile functions.
Inside the transverse area of each pair of colliding nucleons strings
may be taken to be distributed according to the nucleon density of the
Gaussian form.
The number of discs is to be chosen in agreement with the observed
value of the percolation parameter $\rho$ given by Eq. (\ref{rho})
Values of the percolation parameter $\rho$  can be taken from \cite{tarn} for
Au-Au collisions at 62.4 and 200 GeV. They were extracted from the observed
distributions in the transverse momentum at different centralities
and so depend on the impact parameter $b$. They are shown in Fig. \ref{fig1}.
For the LHC energy $E= 2.76 TeV$ we used the conclusions in
\cite{pajares,pajares1} that the values of $\rho$ are roughly 4 times larger than
at RHIC energies.

The number of particpant nucleons for Au-Au collisons at 200 GeV
and given centralities, or equivalently,
$b$ can be borrowed from ~\cite{glau4}. They are reproduced in Table 1.

Taking the radius of both colliding nuclei
$R=A^{1/3}R_0$ with $R_0=1.2$ fm and $r_0=0.3$ fm, for $Au-Au$ collisions
at 200 GeV one gets from Eq. (\ref{rho}) and Fig. \ref{fig1}
the number of strings given at a given $b$. Comparison with Table 1.
then shows that for Au-Au collisions at 200 GeV each pair
of participant nucleons gives rise to approximately 7 colour strings.
Assuming that the number of participant nucleons is purely geometrical
and does not change significantly with energy we conclude that for
Au+Au collisions at 62.4 GeV and 2.76 Tev the numbers of strings
per each pair of participant nucleons are 4 and 28 respectively.

\begin{table}[h]
\begin{center}
\caption{Number of particpant nucleons for given centrality and $b$
for Au+Au at 200 GeV}
\medskip

\begin{tabular}{|c|c|c|}
\hline
Centrality&$b$ (fm)&$N_{part}$\\\hline
  0$-$5   \%&   2.2 &  352.2 \\\hline
  5$-$10  \%&   4.04&  294.7 \\\hline
  10$-$15 \%&   5.27&  245.6 \\\hline
  15$-$20 \%&   6.26&  204.2 \\\hline
  20$-$30 \%&   7.48&  154.5 \\\hline
  30$-$40 \%&   8.81&  103.8 \\\hline
  40$-$50 \%&  10.01&   64.9 \\\hline
  50$-$60 \%&  11.11&   36.6 \\\hline
  60$-$70 \%&  12.06&   18.8 \\\hline
  70$-$80 \%&  12.96&    7.5 \\\hline
  80$-$90 \%&  13.85&    4.4 \\\hline
  \end{tabular}
\end{center}
\end{table}

The strings modeled by discs may overlap in the transverse area and form
clusters of different number $n$ of fused strings and form. Observed
particles are emitted from each cluster with the average multiplicity
\beq
\mu_{nk}=\sqrt{\frac{n\Omega_{nk}}{\Omega_0}}\mu_0
\label{mu}
\eeq
and  average transverse momentum squared
\beq
p^2_{nk}=\sqrt{\frac{n\Omega_0}{\Omega_{nk}}}p_0^2
\eeq
for the $k$-th  cluster of $n$ fused strings. Here
$\Omega_{nk}$ is the transverse area of the cluster
and $\mu_0$ and $p_0^2$ are the
multiplicity and transverse momentum squared for a simple string.
In our picture each particle emitted from a given point in a cluster has
to pass a certain path in the overlap area before being observed.
A part of it has to pass through the same or different clusters and so
looses its energy as described in the previous sections.
During the time of its passage strings partially decay and loose their
colour. This effect is taken into account according to Eq.
(\ref{qtime}). The average length
$l_{nk}$ traveled by the particle emitted from the $k$-th cluster of
$n$ strings depends both on the distribution of clusters and on the direction
of the emission. Due to  azimuthal asymmetry of the cluster distribution
following from the asymmetry of the overlap area the average distribution of
emitted particles will depend on the azimuthal angle and lead to
non-vanishing anisotropic flows.

In the simulation of an event one has, first, to determine the distribution
of clusters in the overlap area and, second, for each cluster to find the
average length $l_{nk}$ of the path which the emitted particle has to travel
inside
the cluster matter. The final distribution of emitted particles in the transverse momentum
and azimuthal angle will be given as a sum
\beq
P(p,\phi)=C\sum_{n,k}\mu_{nk}e^{-\frac{p_0(p,l_{nk})}{\sqrt{T_{nk}}/2}}
\label{ppphi}
\eeq
and the distribution in the azimuthal angle only
\beq
P(\phi)=C\int dp^2P(p,\phi).
\label{pphi}
\eeq
(We use the thermal distribution (\ref{probb}) to be able to move
into the region of $p$ of the order of several GeV/c)

With the number of strings $N$ not very large (below 100) the described
procedure is  realizable on a computer for a reasonable
processing time. However this time grows very fast with the number
of strings (as approximately $\sim N^3$). This motivated a somewhat
simplified approach to our simulations. Instead of taking clusters
at different locations in the transverse space and of different geometric
forms, we assumed them to form a square lattice in the transverse space with
the side length $a=\sqrt{\pi r_0^2}$. Throwing of  discs transforms
these primitive structures into clusters of strings ("ministrings")
with a variable number of fused simple strings equal to the number of disc centers
found in a given lattice cell and thus with a variable cell-dependent
percolation parameter $\rho$. In this manner we avoid the study of
all complicated geometric structures which arise when
a large number of simple strings overlap. Calculations of the momentum
distribution of emitted particles in such a picture shows that it models
the actual cluster formation quite successfully ~\cite{BKPV}.

Note that in Eq. (\ref{ppphi})  it is assumed that
multiplicity $\mu_{nk}$ of the cluster is fixed by its area and colour
according to Eq. (\ref{mu}). This multiplicity
is the average over events. On the event-by-event basis one may take the
multiplicity distributed according to Poisson's law around this average,
as advocated in ~\cite{vechernin}, where it was shown that such distribution
leads to the negative binomial distribution of observed particles.
We shall see that such stochastic emission does not change significantly
the final results for the flow coefficients although naturally somewhat
complicates numerical calculations.

Note that the fluctuations in the distribution of strings
and multiplicities are not the only ones in the dynamics of particle
production. Also the numbers of participant nucleons $N_{part}$
and strings inside
each of them $N_{str}$ fluctuate as well as the values of the impact parameter $b$.
However these additional fluctuations seem to have little to do with the
azimuthal anisotropy, as supported by our results, which show that
fluctuations in the multiplicity do not influence the final anisotropy.
Therefore in our calculations we have not taken into account
fluctuations of $N_{part}$, $N_{str}$ nor $b$, although they can be
essential in other context ~\cite{VK}

In our model, with the coefficient $\alpha$ for the string color decay
fixed at the value 0.03, as explained in the previous section,
only a single  dimensionless parameter $\kappa$ remains,
which characterizes the
loss of energy in passing through the string field and is
to be extracted from the experimental data.
It was adjusted to fit the experimental value of $v_2$
integrated over the transverse momenta for Au-Au collisions
at 200 GeV in mid-central events.

On the event-by-event basis the final flow coefficients $v_n$
are found from the distribution of emitted particles
$P(p,\phi)$ at a given $p$ or
$P(\phi)$, integrated over all $p$ by the standard formulas.
Let in a given event
\beq
a_n=\int d\phi \cos(n\phi)P(\phi),\ \ b_n=\int d\phi \sin(n\phi)P(\phi)
\eeq
then
\beq
v_n=\frac{\sqrt{a^2_n+b^2_n}}{a_0}, n=1,2,....
\eeq

\section{Numerical results}
\begin{figure}
\hspace*{2.5 cm}
\begin{center}
\includegraphics[scale=0.85]{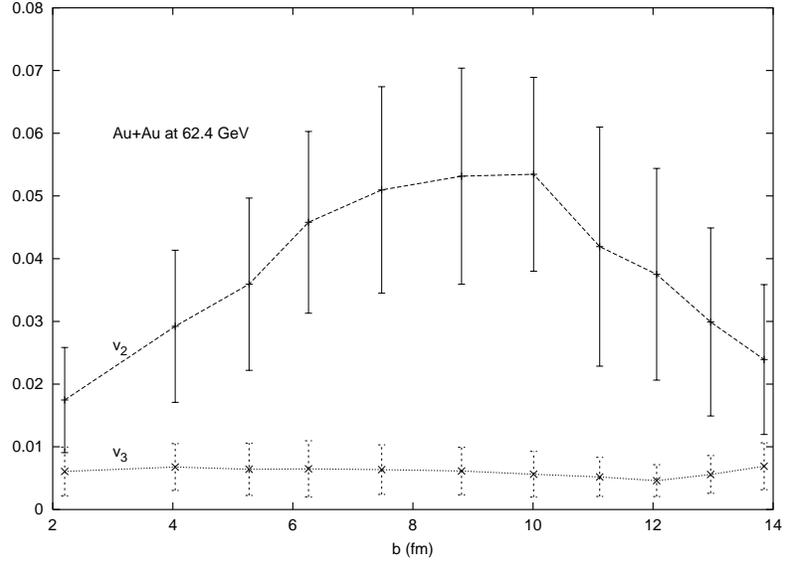}
\end{center}
\caption{$v_2$ and $v_3$ integrated over $p_T$ as a function of $b$ for Au-Au collisions
at $E=62.4$ GeV.}
\label{fig3}
\end{figure}

Our  numerical calculations were performed for
Au-Au collisions  with the profile function
taken according to the Saxon-Woods formula and cluster multiplicities
distributed according to the Poisson law.
The ordinary string tension parameter $T_0$ was taken to agree with the slope
of the spectra in the soft region: $T_0=0.08$ (GeV/c)$^2$
The adjusted value of the quenching coefficient turned out
to be  $\kappa=0.058$.
The number of simulation was taken 100.
Our Monte-Carlo results  obtained for $v_2$ and $v_3$
 integrated
over the transverse momenta and averaged over events are shown as a
function of $b$ in Figs.
\ref{fig3}-\ref{fig5} for Au-Au collisions
at energies 62.4, 200 and 2760 GeV
respectively. In these and the following figures  error bars show
fluctuations of the flow coefficients
around their average values from event to event, not the insufficient
precision of the calculations due to the finite number of simulations.
One observes that these event-by-event fluctuations are
quite large.
\begin{figure}
\hspace*{2.5 cm}
\begin{center}
\includegraphics[scale=0.85]{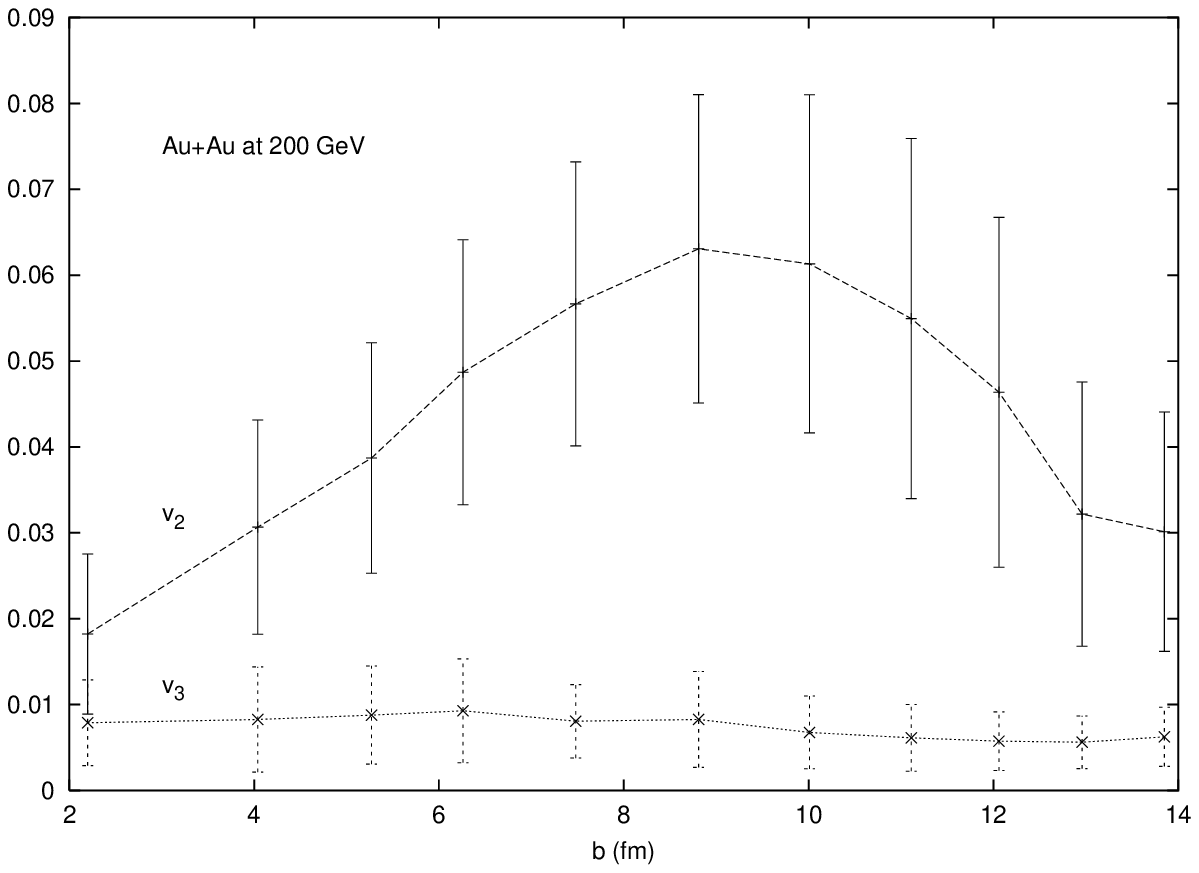}
\end{center}
\caption{Same as for Fig. \ref{fig3} at $E=200$ GeV}
\label{fig4}
\end{figure}
\begin{figure}
\hspace*{2.5 cm}
\begin{center}
\includegraphics[scale=0.85]{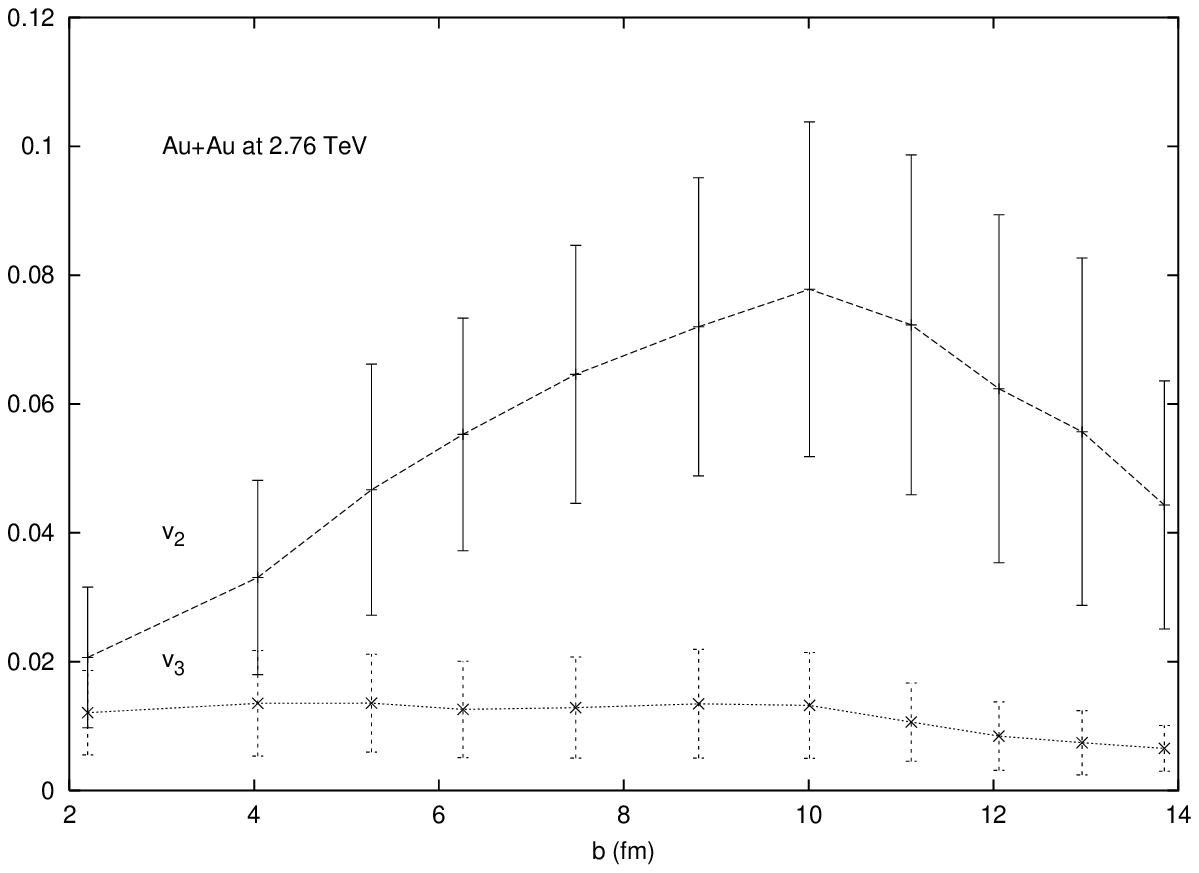}
\end{center}
\caption{Same as for Fig. \ref{fig3} at $E=2.76$ TeV.}
\label{fig5}
\end{figure}

\begin{figure}
\hspace*{2.5 cm}
\begin{center}
\includegraphics[scale=0.85]{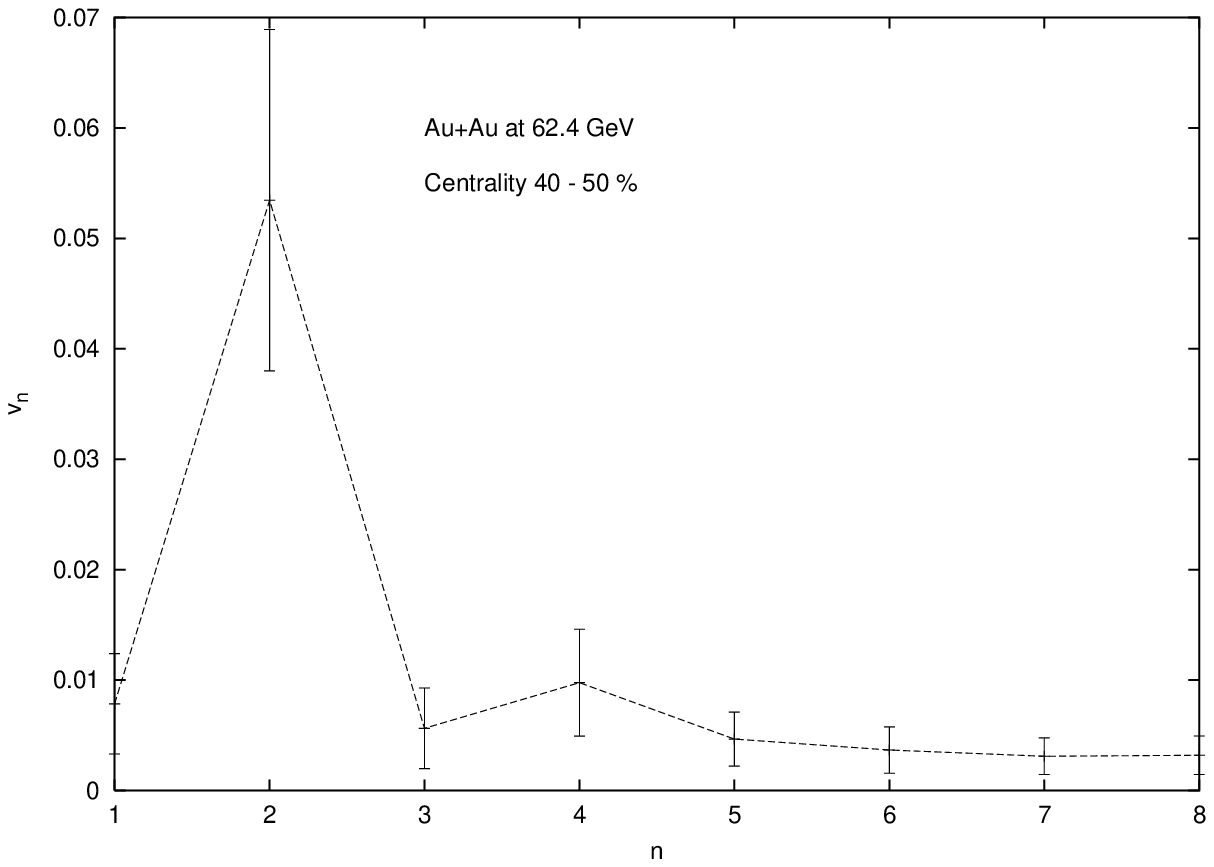}
\end{center}
\caption{$v_n$, $n=1,...8$ integrated over $p_T$ at centrality 40$-$50 \%
for Au+Au collisions at 62.4 GeV}
\label{fig6}
\end{figure}

\begin{figure}
\hspace*{2.5 cm}
\begin{center}
\includegraphics[scale=0.85]{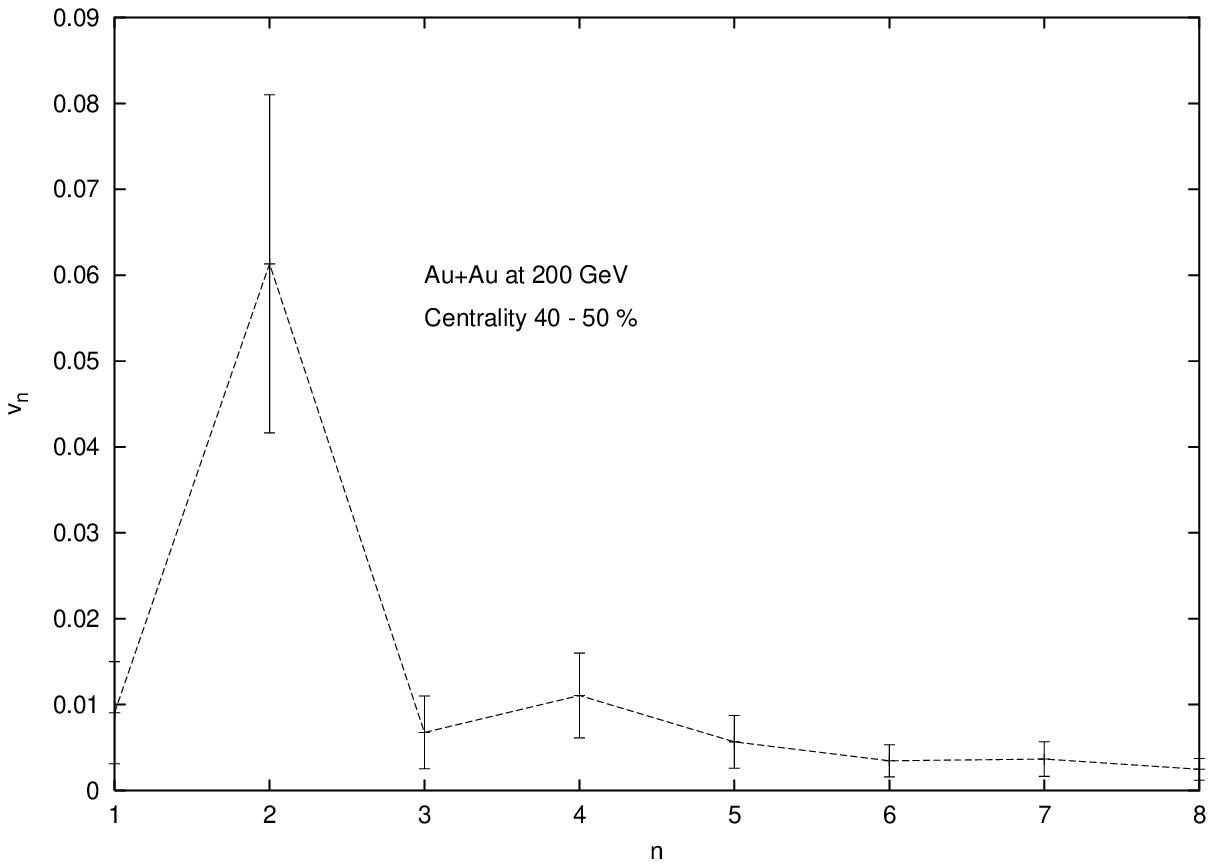}
\end{center}
\caption{Same as for Fig. \ref{fig6} at $E=200$ GeV}
\label{fig7}
\end{figure}

\begin{figure}
\hspace*{2.5 cm}
\begin{center}
\includegraphics[scale=0.85]{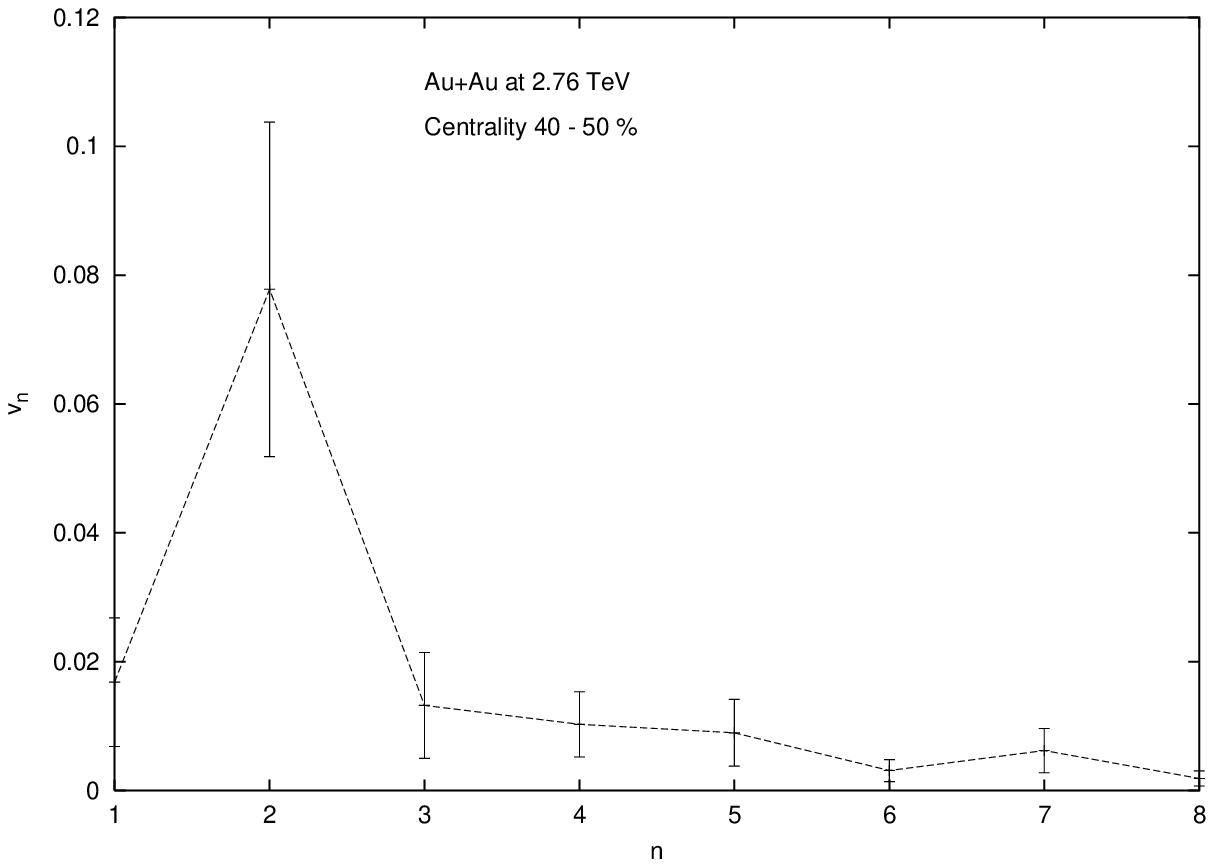}
\end{center}
\caption{Same as for Fig. \ref{fig6} at $E=2.76$ TeV}
\label{fig8}
\end{figure}

To show higher harmonics we present $v_n$ for $n=1,...8$
integrated over $p_T$ at centrality 40$-$50 \% in Figs. \ref{fig6}-
\ref{fig8}, again with their event-by-event fluctuations from the average.

\begin{figure}
\hspace*{2 cm}
\includegraphics[scale=0.45]{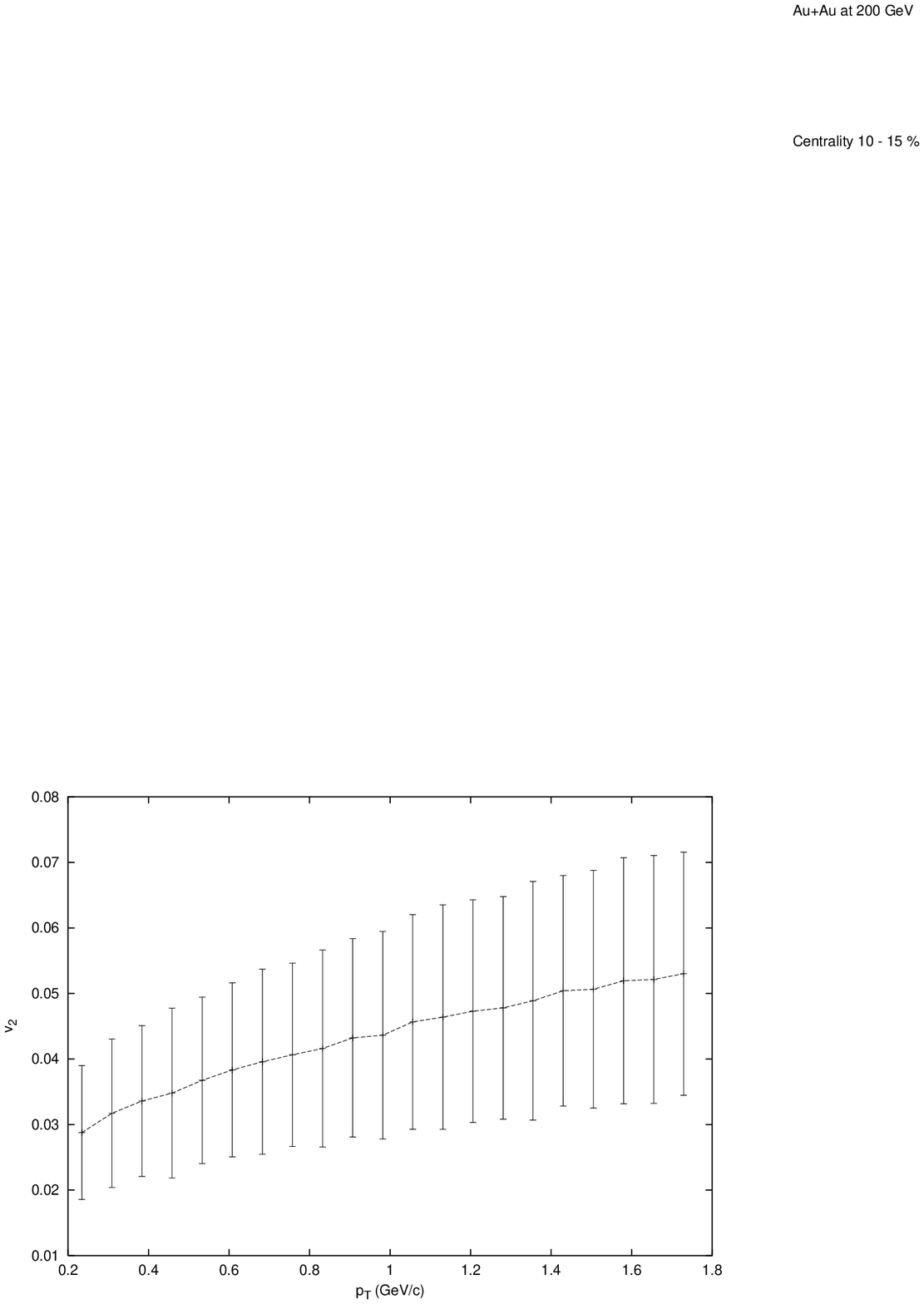}
\vspace*{-3.5 cm}
\hspace*{1 cm}
\includegraphics[scale=0.45]{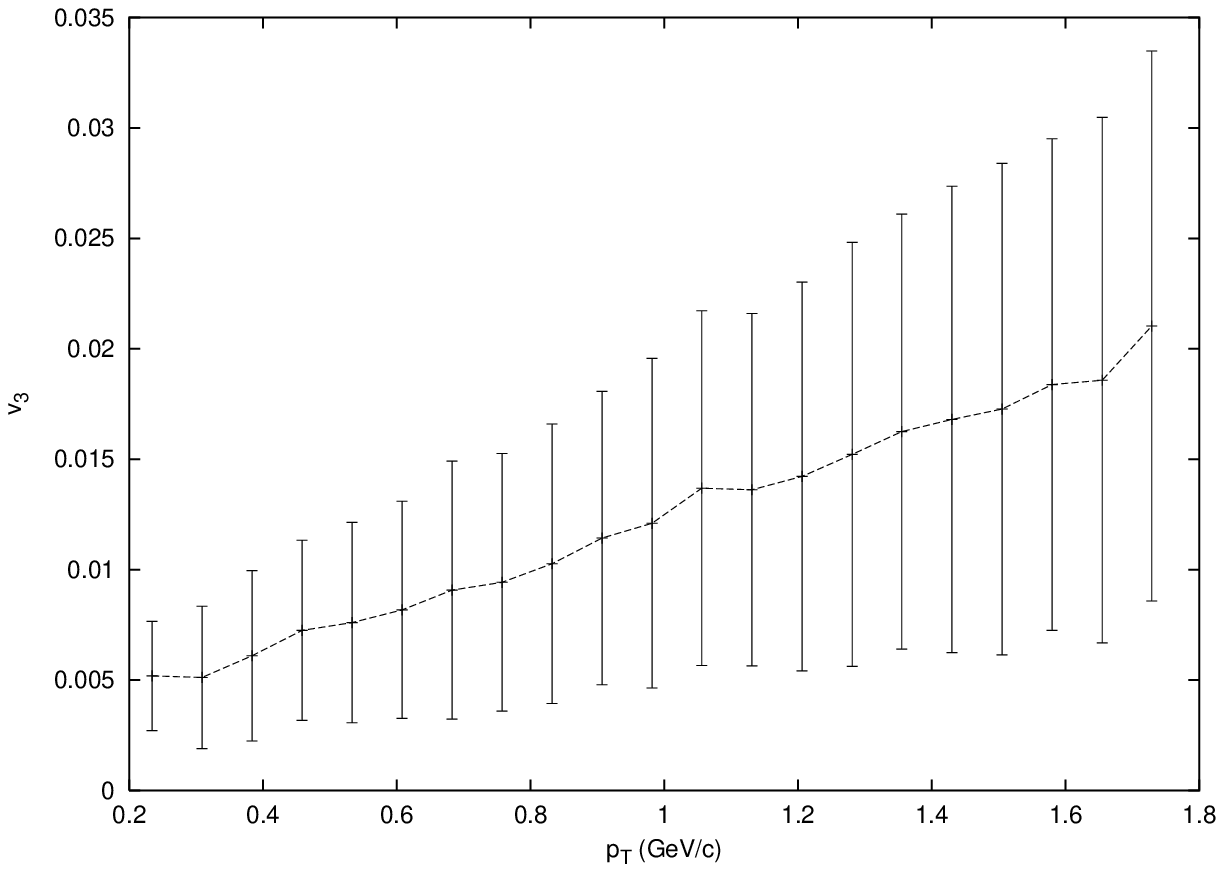}
\vspace*{3.5 cm}
\caption{$v_2$ and $v_3$ as a function of $p_T$ for Au+Au collisions
at 200 GeV and centrality 10$-$15 \%}
\label{fig9}
\end{figure}

\begin{figure}
\hspace*{2 cm}
\includegraphics[scale=0.45]{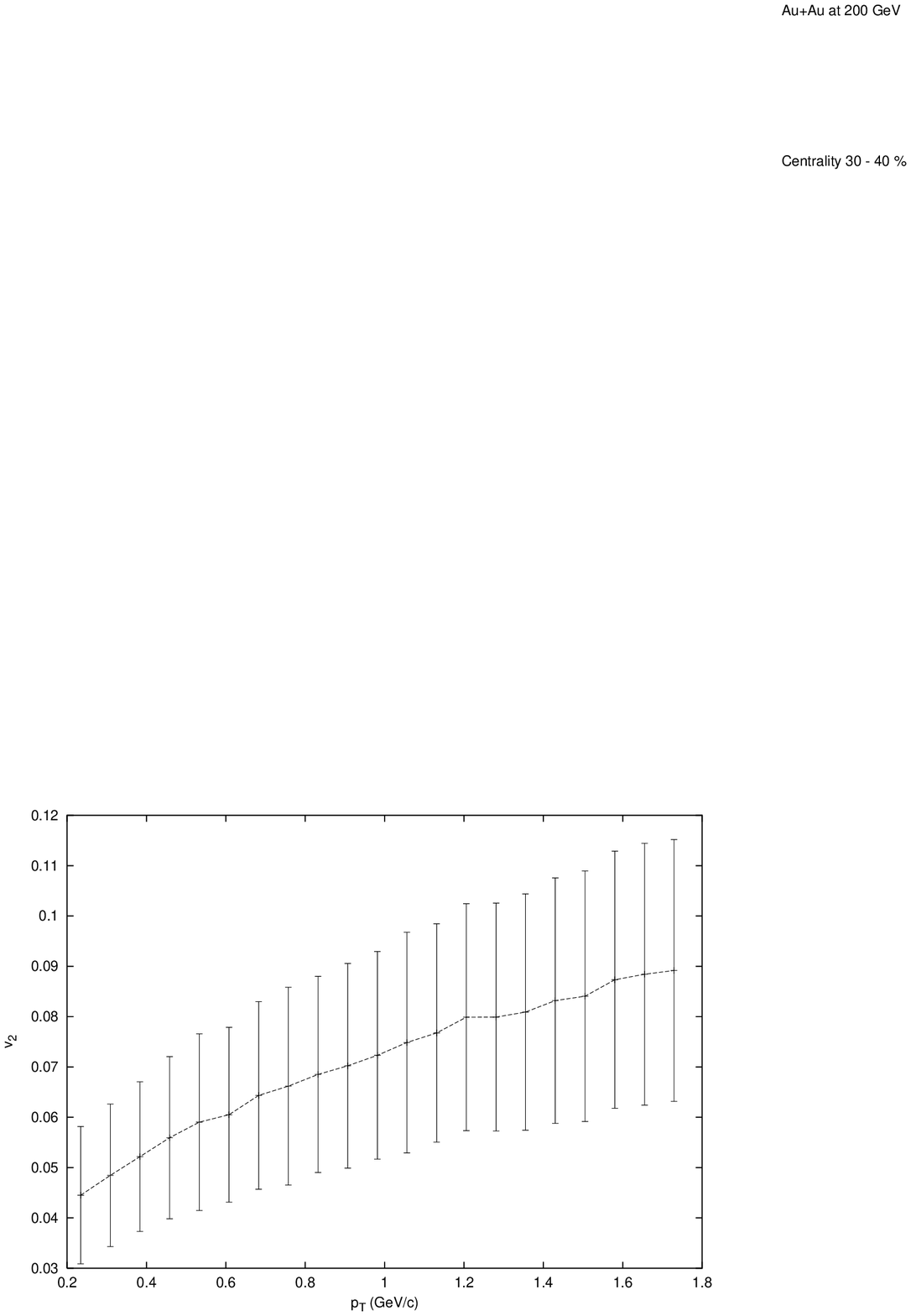}
\vspace*{-3.5 cm}
\hspace*{1 cm}
\includegraphics[scale=0.45]{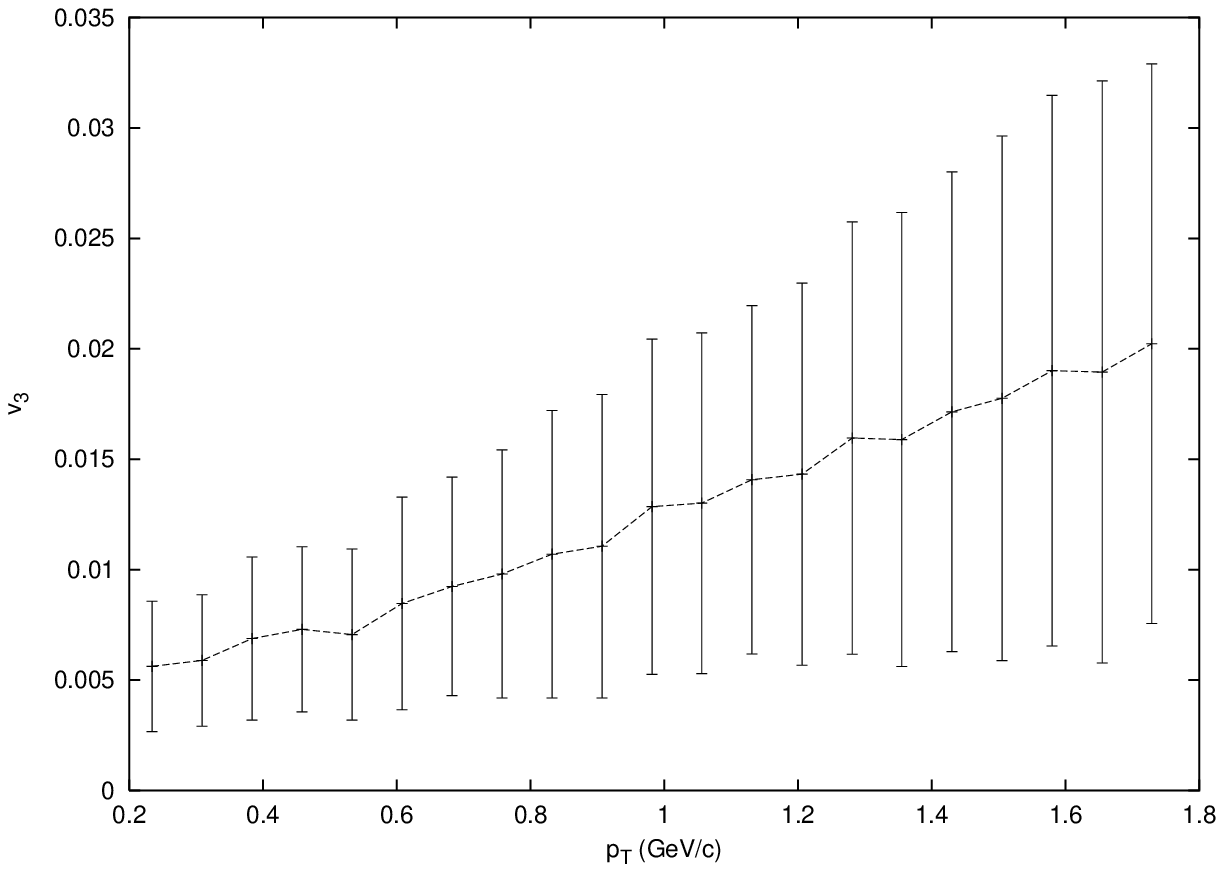}
\vspace{3.5 cm}
\caption{$v_2$ and $v_3$ as a function of $p_T$ for Au+Au collisions
at 200 GeV and centrality 40$-$50 \%}
\label{fig10}
\end{figure}

\begin{figure}
\hspace*{2 cm}
\includegraphics[scale=0.45]{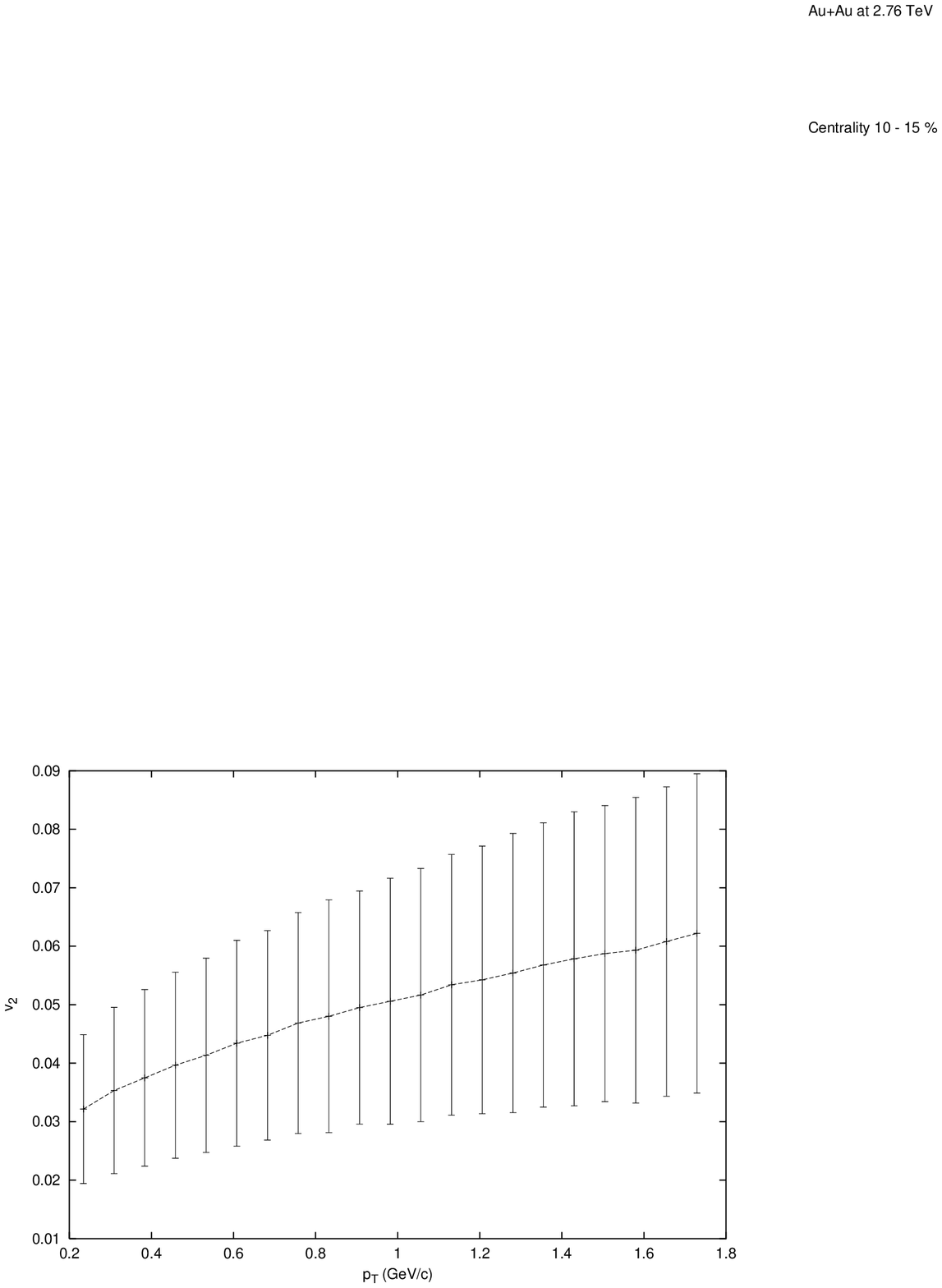}
\vspace*{-3.5 cm}
\hspace*{1 cm}
\includegraphics[scale=0.45]{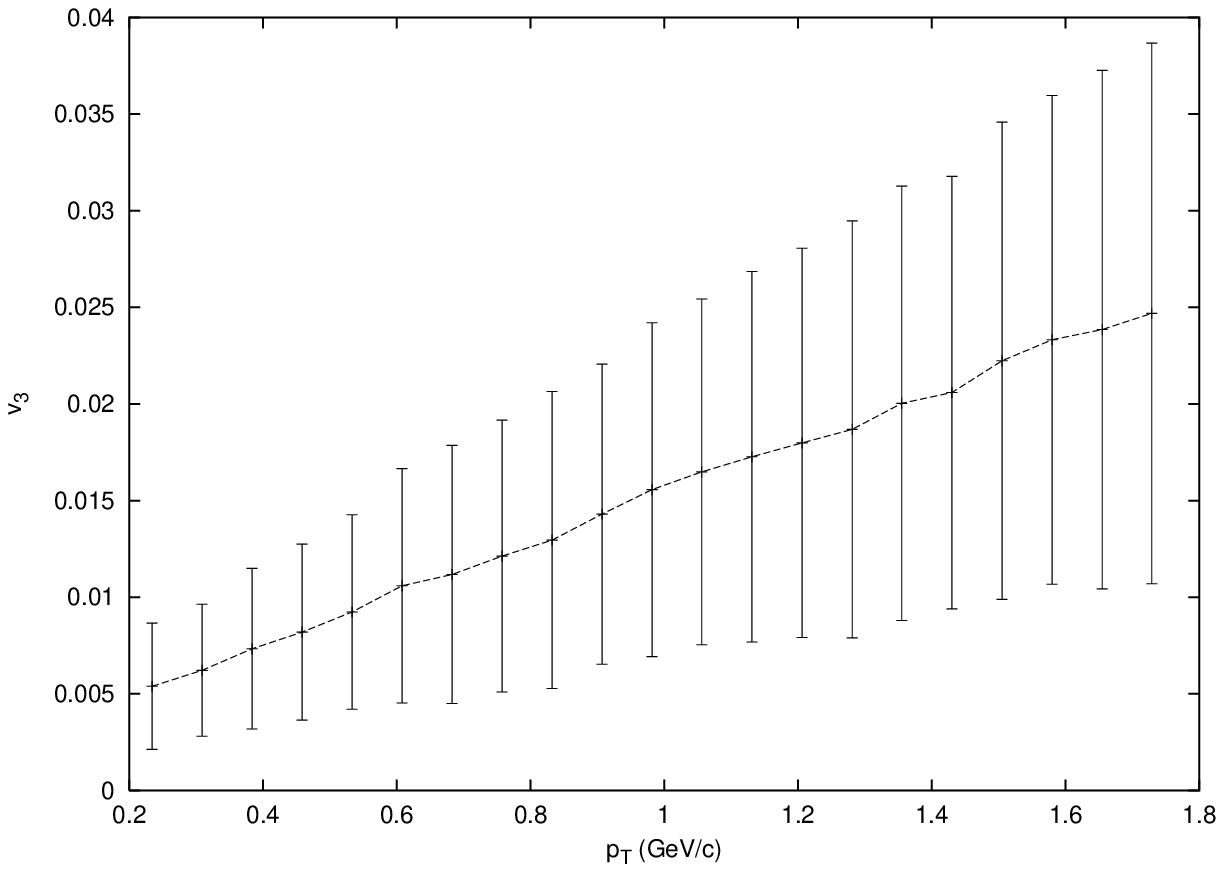}
\vspace{3.5 cm}
\caption{$v_2$ and $v_3$ as a function of $p_T$ for Au+Au collisions
at 2.76 TeV and centrality 10$-$15 \%}
\label{fig11}
\end{figure}

\begin{figure}
\hspace*{2 cm}
\includegraphics[scale=0.45]{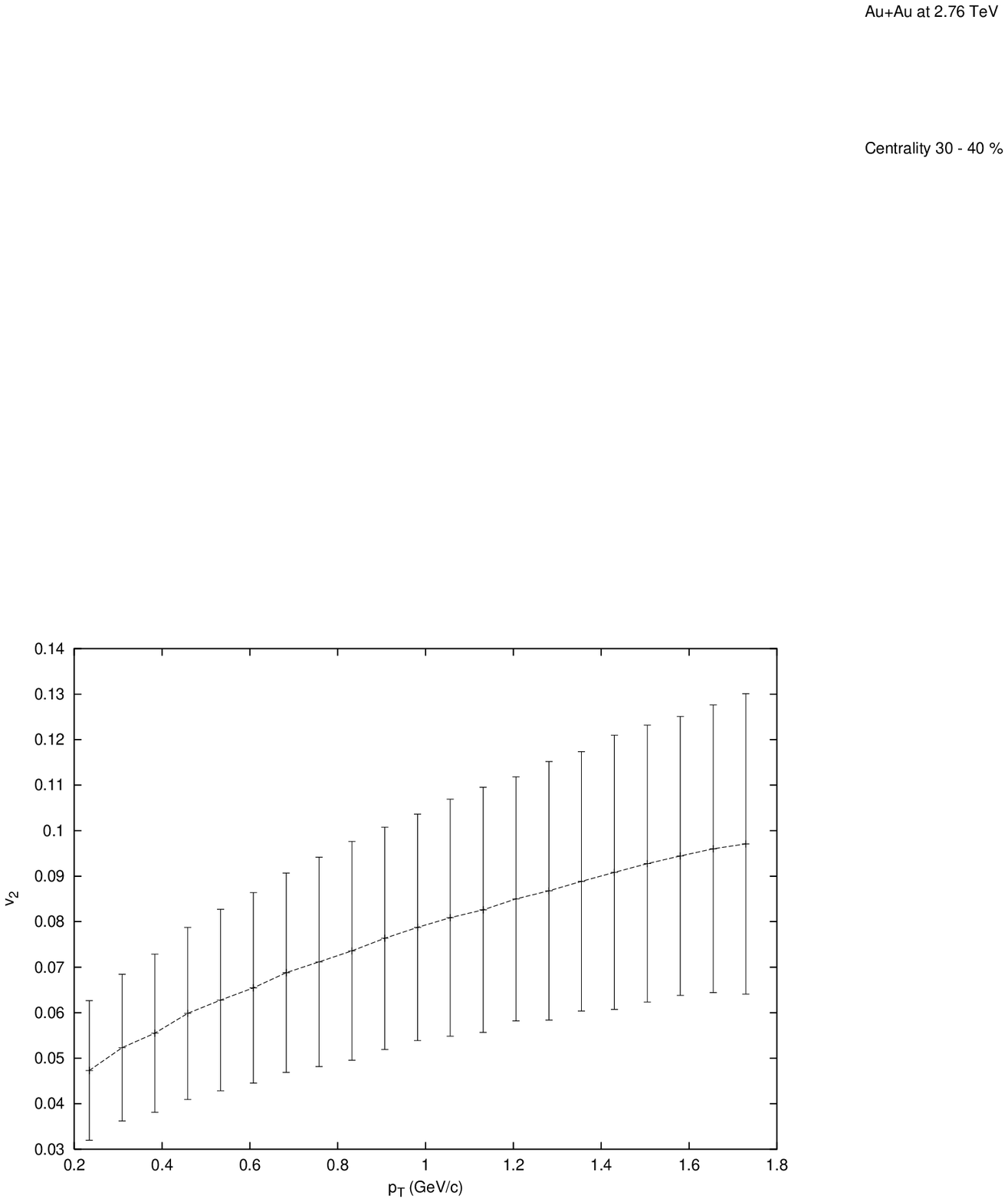}
\vspace*{-3.5 cm}
\hspace*{1 cm}
\includegraphics[scale=0.45]{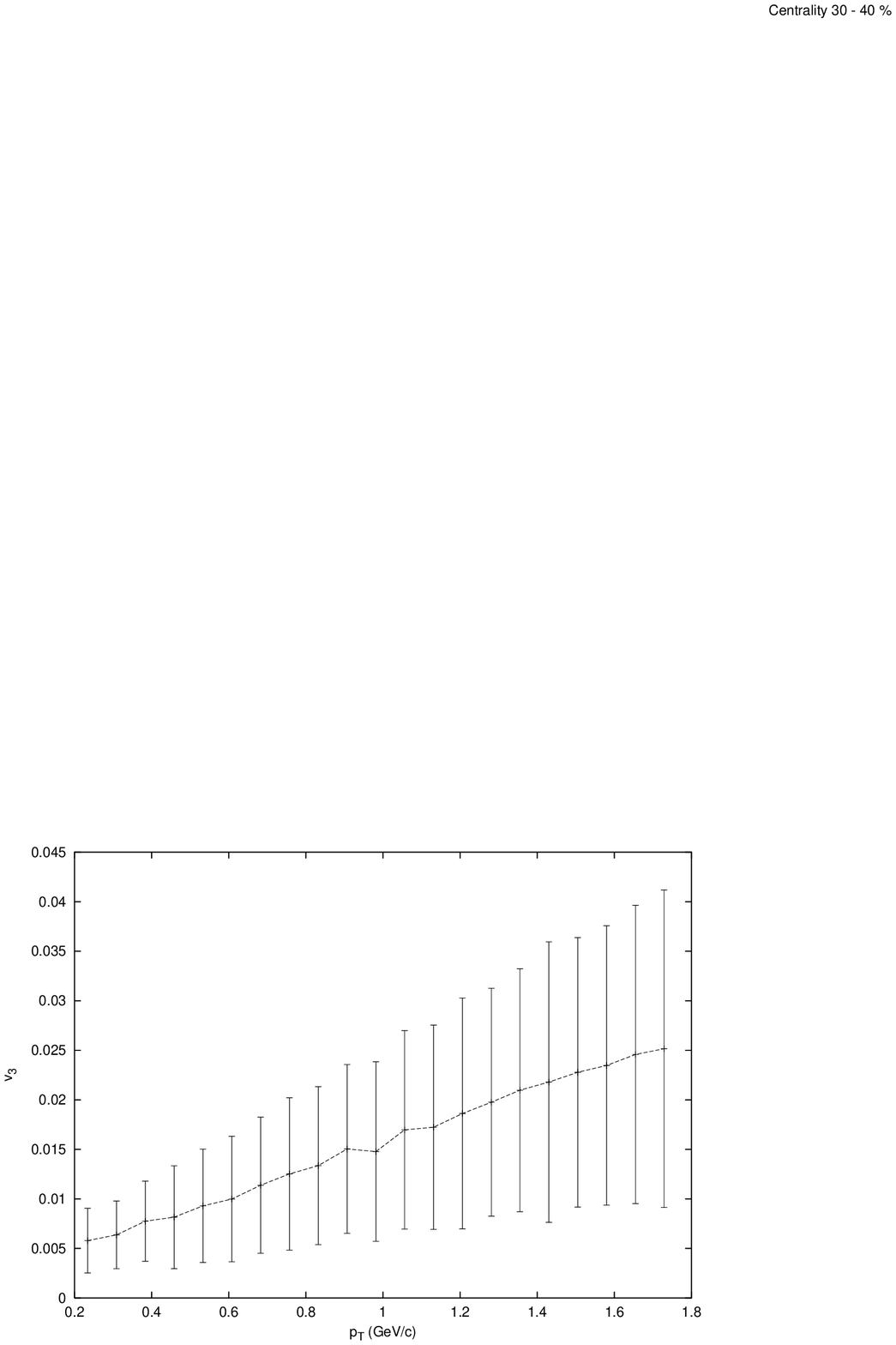}
\vspace{3.5 cm}
\caption{$v_2$ and $v_3$ as a function of $p_T$ for Au+Au collisions
at 2.76 TeV and centrality 40$-$50 \%}
\label{fig12}
\end{figure}

Finally in Figs. \ref{fig9} -\ref{fig12}
we present our results for the $p_T$ dependence of $v_2$ and $v_3$
at two centralities 10$-$15 \% and 40$-$50 \% for Au+Au collisions
at energies 200 GeV and 2.76 TeV

Inspecting these results we see that the $b$- and $p_T$ dependence
of $v_2$ as well as their absolute values well agree with the experimental
data (which of course is based on the proper adjustment of coefficient
$\kappa_0$ at $E=200$ GeV and medium centrality). The behaviour of the
triangular
coefficient $v_3$ is also found  to agree with the experimental observations.
Its average values are calculated to be somewhat smaller than the
experimental values. However their event-by-event
fluctuations from the average turn out to be very large,
so that taking them into account may put the experimental values
well within the calculated ones.

To analyze the important components of fluctuations leading to relatively
large values of $v_3$ we repeated our calculations
substituting Poisson distributed cluster
multiplicities by their average. For Au+Au collisions at 200 GeV and
centrality 40$-$50 \% this gives values for $v_2$ and $v_3$ integrated over
$p_T$   0.594E-01 $\pm$ 0.145E-01 and  0.574E-02$\pm$  0.403E-02
respectively to be compared with  0.588E-01 $\pm$ 0.186E-01
and  0.668E-02 $\pm$  0.399E-02 with Poisson distributed multiplicities.
As one observes the difference is insignificant.
On the other hand the importance of quenching is found to be overwhelming.
Taking $\kappa_0=0$ but keeping the Poisson distributed multiplicities
one obtains values for $v_2$ and $v_3$    0.180E-02$\pm$  0.839E-03
and  0.112E-02$\pm$  0.622E-03, that is practically zero.

\section{Conclusions}
We have performed detailed Monte-Carlo simulations to study
anisotropic flows in the percolating string scenario.
We have confirmed that the colour string model with fusion and
percolation can successfully describe the observed elliptic and triangular
flows in high-energy heavy-ion collisions. An important ingredient in
this description is  anisotropy of the string
emission spectra in the azimuthal direction which follows
from quenching of the emitted partons in the strong colour field
inside the string. As a mechanism for this quenching we used the radiation
energy loss during propagation of a fast charged particle
in a constant field, borrowed from the QED.
Upon adjusting the parameter of
quenching, this allowed to describe the data quite well both in their energy
centrality and transverse momentum dependence.
We have also studied higher anisotropic flow coefficients from $v_4$ to
$v_8$ which are found to be small as compared to  $v_2$.

To compare with our earlier calculations in the grossly oversimplified
picture in ~\cite{brapaj1} we have found more pronounced energy dependence.
This follows from the explicit dependence of the
quenching in the QED on the field strength, which is translated into
dependence on the string tension in the colour string picture.
So, in contrast to our phenomenological formula adopted in ~\cite{brapaj1},
this quenching is not purely geometrical but grows with the percolation
parameter even when all strings are fused into a single cluster occupying
the whole overlap area.

Our results are based on strings which have infinite dimensions in rapidity.
So from the start they refer to the central region of nearly zero rapidity
and thus do not allow to study the rapidity dependence of the flow
coefficients. To do this we have to introduce strings of finite rapidity
length and in this way take into account energy conservation. This
complicates our picture substantially and will be the object of our
further studies.

\section{Acknowledgements}
The authors are greatly thankful to N.Armesto for his help in the
technique of distributing nucleons and strings in the transverse space
and constructive comments.
This work is done under the projects FPA2008-01177 and Consolider of the
Ministry of Science and Innovation of Spain and under the project of
Xunta de Galicia. One of the autors (M.A.B) is indebted to the
University of Santiago de Compostela for attention and financial support.
M.A.B. and V.V.V. have also benefited from
grant RFFI 12-02-00356-a which partially supported this work.


\end{document}